\documentstyle[11pt]{article}
\begin{document} 
\makeatletter
\@addtoreset{equation}{section}
\makeatother
\renewcommand{\theequation}{\thesection.\arabic{equation}}
\baselineskip 15pt

\title{\bf Reply to ``Comment on: A Quantum Approach to Static Games of
 Complete Information''
\footnote{Work supported in part by Istituto Nazionale di Fisica Nucleare,
 Sezione di Trieste, Italy}}
\author{Luca Marinatto\footnote{e-mail: marinatto@ts.infn.it}\\
{\small Department of Theoretical Physics of the University of Trieste, and}\\
and \\
\\ Tullio Weber\footnote{e-mail: weber@ts.infn.it }\\
{\small Department of Theoretical Physics of the University of Trieste, and}\\
{\small Istituto Nazionale di Fisica Nucleare, Sezione di Trieste, Italy.}}
\date{}

\maketitle

%-----------------------------------------------------------------------------

In a recent comment \cite{ref1} S.C. Benjamin made some interesting remarks
 about the terminology, the postulates and the mathematical results of a new
 quantum scheme \cite{ref2} proposed in order to solve the paradigmatic
 dilemma arising in a famous game ($\,${\em Battle of the Sexes}$\,$) in the
 framework of the classical Theory of Games. 
In answering these remarks we take the opportunity of outlining some topical
 points of our work which we hold to be crucial for a better comprehension of
 our new approach to the Quantum Theory of Games, and which differs from
 the other proposed versions \cite{ref3} and \cite{ref4}.

As regards terminology, we think that the choice of calling {\em strategies}
 the quantum states instead of the operators used to manipulate them, is very
 natural and quite consistent with the spirit of the
 classical game theory.
In fact, in the classical framework of the theory, starting a game each player
 has at his disposal an ensemble of {\em strategies} ${\cal{S}}$ which is no
 more than a set ($\,$i.e. a collection of objects, equipped with the only
 concept of {\em ``belonging to''} represented by the symbol $\in$$\,$). 
In pursuing the aim of finding new results beyond the ones which are predicted
 by the classical theory, we felt as a natural request adding an Hilbert space
 structure to this strategic set ${\cal{S}}$, in order to be allowed to use the
 methods of Quantum Mechanics.
As a consequence the pure strategies of the classical framework are now
 represented by an orthonormal and complete set of vectors, and their linear
 combinations
 represent the mixed ($\,$i.e. probabilistic$\,$) strategies,
 the probabilities being quantified by the squared modula of the complex
 coefficients of the combination.
Clearly this strategic quantum space exhibits a richer structure, since it
 contains not only the subset of the factorizable states, which correspond to
 the pure and mixed strategies of the classical theory, but also
 non-factorizable states, which we hold to constitute the real powerful tool
 enabling the players to solve the unescapable dilemma of the classical theory.
On the other hand, in our terminology the act of choosing a move ($\,$which for
 Eisert et al. \cite{ref3} corresponds to a {\em strategy}$\,$) means choosing
 in a given set of operators the one to be applied to the quantum strategies,
 and does not belong to the strategic space, pertaining instead to the
 cathegory of {\em tactics} which can be used in that space.

The second issue faced in the comment concerns the choice of the {\em
 tactics set}, i.e. the set of local operations each player has at his disposal
 in order to manipulate his own starting strategy.
Our main purpose was consisting in choosing the smallest set of operations
 able to reproduce, when applied to a factorizable couple of
 strategies, the results of the classical Theory of Games.
In fact we have been able to show that, limiting ourselves to a probabilistic
 choice
 between operators $I$ and $\sigma_{x}$, was sufficient to reach our goal.
On the other hand, our belief was that new results should have come out from
the richer structure of the strategic space, i.e. from taking into account
entangled couple of strategies, and in fact it happened.
Obviously the class of allowed manipulations could be enlarged, but in our
 paper we did not take care of this possibility since
 our minimal choice was enough to reproduce intact the classical results when
 considering only factorizable strategies and to obtain the disappearance of
 the dilemma when resorting to entangled strategies.
Therefore we do not reject the possibility of improving our scheme of the
 Quantum Theory of Games ($\,$for static games and complete information$\,$)
by extending the set of tactics to include other quantum mechanical
 manipulations, provided that the results of the classical behaviour are
 always reproduced.

The third remark, concerning the impossibility of solving the dilemma even
 when resorting to the entangled strategy $\vert \psi \rangle =1/{\sqrt{2}}(\,
 \vert OO \rangle + \vert TT \rangle \,)$ is the most important one.
The author of the comment claims that the dilemma persists since the players
 cannot decide between manipulating the initial strategy
 $(p^{\star}=0,q^{\star}=0)$ and leaving it unchanged
 $(p^{\star}=1,q^{\star}=1)$.
We maintain that the claim is not correct, since both the choices will
 eventually lead to the same final strategy, which is exactly the same the
 players possess from the beginning of the game. It is therefore apparent, on
 the ground of reasonableness, that both players, knowing this fact, will
 decide to do nothing on their strategies. 
Both players are in fact restrained from manipulating their own strategy since
 there exist at least one possibility of obtaining the wrong result ($\,$this
 will happen for example if Alice decides to choose $p^{\star}=0$ while Bob
 chooses $q^{\star}=1$$\,$).
The choice wheter manipulating or not the compound strategy
 $\vert \psi \rangle$ is not a matter of cost ($\,$in fact {\em doing nothing}
 could be considered cheaper than {\em doing something}, in terms of resources
 needed to operate on the strategies$\,$), but instead it represents
 the most rational behaviour of the two players.
Which player will in fact decide to refuse a certain gain, without prospects of
 a better gain and running the risk of incurring a loss?
The hypothesis of Complete Information and the supposed rational way of 
thinking of the players forces us to affirm again that our quantum approach to
 {\em Battle of the Sexes} game is able to solve the
 dilemma of the existence of two equally attractive Nash strategies which is
 present in its classical version.

%---------------------------------------------------------------------

\end{document}